\documentclass[twocolumn,aps,pra,superscriptaddress]{revtex4}
\usepackage[latin9]{inputenc}
\usepackage{bm}
\usepackage{amsmath}
\usepackage{amssymb}
\usepackage{esint}

\makeatletter
\@ifundefined{textcolor}{}
{%
 \definecolor{BLACK}{gray}{0}
 \definecolor{WHITE}{gray}{1}
 \definecolor{RED}{rgb}{1,0,0}
 \definecolor{GREEN}{rgb}{0,1,0}
 \definecolor{BLUE}{rgb}{0,0,1}
 \definecolor{CYAN}{cmyk}{1,0,0,0}
 \definecolor{MAGENTA}{cmyk}{0,1,0,0}
 \definecolor{YELLOW}{cmyk}{0,0,1,0}
}

\usepackage{color}
\usepackage{braket}

\usepackage{subfigure}\usepackage{epsfig}\usepackage{amsfonts}\usepackage{mathrsfs}\usepackage{color}\usepackage[toc,page,title,titletoc,header]{appendix}

\makeatother

\begin{document}
%\preprint{ZJNU\&RUC}


\title{Universal Relations of an Ultracold Fermi Gas with Arbitrary Spin-Orbit
Coupling}

\title{Universal Relations of an Ultracold Fermi Gas with Arbitrary Spin-Orbit
Coupling}
\affiliation{Department of Physics, Renmin University of China, Beijing, 100872,
China}

\affiliation{Beijing Computational Science Research Center, Beijing, 100084, China}

\affiliation{Beijing Key Laboratory of Opto-electronic Functional Materials \&
Micro-nano Devices (Renmin University of China)}
\author{Jianwen Jie}

\author{Ran Qi}
\email{qiran@ruc.edu.cn}
\affiliation{Department of Physics, Renmin University of China, Beijing, 100872,
China}
\affiliation{Beijing Key Laboratory of Opto-electronic Functional Materials \&
Micro-nano Devices (Renmin University of China)}
\author{Peng Zhang$^{1,2,3}$}
\email{pengzhang@ruc.edu.cn}\date{today}

%Abstract
\begin{abstract}
We derive the universal relations for an ultracold two-component Fermi
gas with an spin-orbit coupling (SOC) $\sum_{\alpha,\beta=x,y,z}\lambda_{\alpha\beta}\sigma_{\alpha}p_{\beta}$,
where $p_{x,y,z}$ and $\sigma_{x,y,z}$ are the single-atom momentum
and Pauli operators for pseudo spin, respectively, and the SOC intensity
$\lambda_{\alpha\beta}$ could take arbitrary value. We consider the
system with an $s$-wave short-range interspecies interaction, and
ignore the SOC-induced modification for the value of the scattering
length. Using the first-quantized approach developed by S. Tan (Phys.
Rev. Lett. \textbf{107}, 145302 (2011)), we obtain the short-range
and high-momentum expansions for the one-body real-space correlation
function and momentum distribution function, respectively. For our
system these functions are $2\times2$ matrix in the pseudo-spin basis.
We find that the leading-order ($1/k^{4}$) behavior of the diagonal
elements of the momentum distribution function (i.e., $n_{\uparrow\uparrow}({\bf k})$
and $n_{\downarrow\downarrow}({\bf k})$) are not modified by the
SOC. However, the SOC can significantly modify
the large-$k$ behaviors of the distribution difference $\delta n({\bf k})\equiv n_{\uparrow\uparrow}({\bf k})-n_{\downarrow\downarrow}({\bf k})$ as well as the
\textit{non-diagonal elements} of the momentum distribution function,
i.e., $n_{\uparrow\downarrow}({\bf k})$ and $n_{\downarrow\uparrow}({\bf k})$. In the absence of the SOC, the leading order
of $\delta n({\bf k})$, $n_{\uparrow\downarrow}({\bf k})$ and $n_{\downarrow\uparrow}({\bf k})$ are ${\cal O}(1/k^{6})$. When SOC appears, it can
induce a term on the order of $1/k^{5}$ for these elements. We further derive the adiabatic
relation and the energy functional. Our results show the SOC can induce
a new term in the energy functional, which describe the contribution
from the SOC to the total energy. In addition, the form of the adiabatic
relation for our system is not modified by the SOC. Our results are
applicable for the systems with any type of single-atom trapping potential,
which could be either diagonal or non-diagonal in the pseudo-spin
basis.
\end{abstract}
\maketitle

\section{Introduction}

In the recent decade a number of new kinds of experimental methods,
such as magnetic and optical Feshbach resonances, were invented to
tune the inter-atomic interactions for alkali and alkaline-earth atoms
\cite{Feshbach,Feshbach2}. A vast number of efforts were devoted
to investigate the strongly interacting state of matter in such systems
\cite{manybody1,manybody2}. The existence of strong correlation effects
and the lack of small parameters for perturbative calculations make
it very difficult to obtain any rigorous results for such strongly
interacting systems. In recent years, a series of exact universal
relations were established which for the first time set up a bridge
between the microscopic short distance correlations and the macroscopic
thermodynamic characters of the system \cite{contact,contactb,contactc,contact2,contact2b,contact3,contact4,contact5,contact6,contact7,contact7b,TanPRL2011,CastinPRA2012,contact8,contact9,contact10,contact11,contact12,contact13,contact14,contact15,contact16}.
These relations show that many important properties of the system
(e.g., the one-body momentum distribution, one-body spatial correlation
function and many-body total energy) are related together via a parameter
which contains the information of the interaction effect in the high-momentum
limit, and is usually called as contact. These relations has already
been confirmed in several different experiments \cite{contact_exp1,contact_exp2,contact_exp3,contact_exp4}.

Another recent important progress in the research of ultracold gases
is the successful experimental realization of synthetic coupling between
atomic (pseudo) spin and momentum \cite{soc exp1,soc exp1.1,soc exp2,soc exp3,soc2d1,soc2d2}.
Here we call this coupling as spin-orbit coupling (SOC). In these
systems the SOC can strongly affect the one-body dispersion and thus
seriously change the many-body properties. Therefore, the SO-coupled
ultracold gases have attracted many attentions from both theorists
and experimentalists \cite{Xiaoling soc,Peng soc1,Peng soc2,Hui review,Yuzhenhua1,Yuzhenhua2}.

Thus, it is very natural to consider the universal relations in the
SO-coupled ultracold gases. Recently S. Peng \textit{et. al.} studied
this problem for a system with a three-dimensional isotropic SOC (${\bm{\sigma}}\cdot{\bf p}$-type)
\cite{PengshiguoPRL2017}. Here we derive the universal relations
for the ultracold gases with arbitrary type of SOC. Explicitly, we
consider an ultracold gas of identical Fermi atoms with pseudo spin
$\uparrow$ and $\downarrow$, with a general type SOC which can be
expressed as $\sum_{\alpha,\beta=x,y,z}\lambda_{\alpha\beta}\sigma_{\alpha}p_{\beta}$,
where $p_{x,y,z}$ are the components of single-atom momentum and
the SOC intensity $\lambda_{\alpha\beta}$ could be arbitrary real
numbers. In our system there could also be a general single-atom trapping
potential, which can be expressed as an arbitrary atomic-position-dependent
$2\times2$ matrix in the pseudo-spin basis.

We further assume that there is an $s$-wave short-range interaction
between two atoms in different pseudo-spin states, which is described
by the scattering length $a$. In principle, the SOC may modify the
value of $a$ \cite{Peng soc1,PengshiguoPRL2017}. Nevertheless, previous
analytical \cite{Peng soc1} and numerical studies \cite{Xiaoling soc}
have shown that in many ultracold gases, including the current experimental
systems with Raman-beam-induced one-dimensional SOC, the SOC-induced
modification of scattering length is negligible when the characteristic
length of the SOC (defined as the inverse of the SOC intensity) is
much larger than the range of the two-body interaction
(i.e., the van der Waals length).  This is normal situation in cold atom system. Thus, the
starting point of our paper is that the scattering length is not changed
by the SOC. This starting point is very different from the assumption
in the recent work by S. Peng \textit{et. al.} \cite{PengshiguoPRL2017}.

\subsection{Our main results}

For the convenience of the readers, here we briefly summarize
the results we obtained.

\textbf{(A)} We derive the short-range expansion of the single-atom
spatial correlation function $\rho_{\sigma\sigma^{\prime}}({\bf r},{\bf r}+{\bf b})$
($\sigma,\sigma^{\prime}=\uparrow,\downarrow$) of a many-body state
$|\Psi\rangle$, which is defined as
\begin{equation}
\rho_{\sigma\sigma^{\prime}}({\bf r},{\bf r}+{\bf b})\equiv\langle\Psi|\hat{d}_{\sigma}^{\dagger}(\textbf{r})\hat{d}_{\sigma'}(\textbf{r}+\textbf{b})|\Psi\rangle\label{rhorr}
\end{equation}
in the second quantized language and can form a $2\times2$ matrix
in the pseudo-spin basis. Here $\hat{d}_{\sigma}^{\dagger}(\textbf{r})$
and $\hat{d}_{\sigma}(\textbf{r})$ are the creation and annihilation
operators of an atom with pseudo spin $\sigma$ at position ${\bf r}$.
For the case where $b\equiv|{\bf b}|$ is small, we expand $\rho_{\sigma\sigma^{\prime}}({\bf r},{\bf r}+{\bf b})$
up to the 2nd order of ${\bf b}$ (see Eq. (\ref{rhor})). We find
that the behavior of $\rho_{\sigma\sigma^{\prime}}({\bf r},{\bf r}+{\bf b})$
is not changed by the SOC in the 0th and $1$st order of ${\bf b}$.
Nevertheless, in the $2$nd order of ${\bf b}$ a new non-analytical
term can be induced by the SOC. This term is the leading non-analytical
term of the non-diagonal elements $\rho_{\downarrow\uparrow}({\bf r},{\bf r}+{\bf b})$
and $\rho_{\uparrow\downarrow}({\bf r},{\bf r}+{\bf b})$.

\textbf{(B)} We derive the high-momentum expansion of the single-atom
momentum distribution function $n_{\sigma\sigma^{\prime}}({\bf k})$
($\sigma,\sigma^{\prime}=\uparrow,\downarrow$) of a many-body state
$|\Psi\rangle$, which is defined as
\begin{equation}
n_{\sigma\sigma^{\prime}}({\bf k})\equiv(2\pi)^{3}\langle\Psi|\hat{d}_{\sigma}^{\dagger}(\textbf{k})\hat{d}_{\sigma'}({\bf k})|\Psi\rangle,\label{rhokk}
\end{equation}
with $\hat{d}_{\sigma}^{\dagger}(\textbf{k})$ and $\hat{d}_{\sigma}(\textbf{k})$
being the creation and annihilation operators of an atom with pseudo
spin $\sigma$ and in the plane-wave state $|{\bf k}\rangle$ defined
by $\langle{\bf r}|{\bf k}\rangle=e^{i{\bf k}\cdot{\bf r}}/(2\pi)^{\frac{3}{2}}$
\cite{norm}. It is clear that $n_{\sigma\sigma^{\prime}}({\bf k})$
can also form a $2\times2$ matrix in the pseudo-spin basis. In previous
researches people mainly study the behavior of the diagonal elements
$n_{\uparrow\uparrow}({\bf k})$ and $n_{\downarrow\downarrow}({\bf k})$.
For our system the non-diagonal terms $n_{\uparrow\downarrow}({\bf k})$
and $n_{\downarrow\uparrow}({\bf k})$ are also non-zero and should
be studied. In the large-$k$ limit we expand all the four elements
$n_{\sigma\sigma^{\prime}}({\bf k})$ to the order of $1/k^{5}$ (see
Eq. ({\ref{rhokexpan}}) and (\ref{rhoud})), and find that:
\begin{itemize}
\item The leading-order ($1/k^{4}$) behavior of the diagonal elements $n_{\uparrow\uparrow}({\bf k})$
and $n_{\downarrow\downarrow}({\bf k})$ are not modified by the SOC.
No matter if there is an SOC or not, we always have $\lim_{k\rightarrow\infty}n_{\uparrow\uparrow}({\bf k})=\lim_{k\rightarrow\infty}n_{\downarrow\downarrow}({\bf k})=C/k^{4}$.
Here the parameter $C$ can be defined as the contact corresponding
to the state $|\Psi\rangle$. As in the systems without SOC, the value
of $C$ is also related to the small-$b$ behavior of $\rho_{\sigma\sigma^{\prime}}({\bf r},{\bf r}+{\bf b})$.
\item The SOC can modify the behavior of $n_{\uparrow\uparrow}({\bf k})$
and $n_{\downarrow\downarrow}({\bf k})$ in the sub-leading order
$(1/k^{5})$ by inducing a new term which is proportional to the contact
$C$. However, the behavior of the total momentum distribution $n({\bf k})\equiv n_{\uparrow\uparrow}({\bf k})+n_{\downarrow\downarrow}({\bf k})$
is still not modified by the SOC on this order.

\item   In the absence of the SOC, the difference $\delta n({\bf k})\equiv n_{\uparrow\uparrow}({\bf k})-n_{\downarrow\downarrow}({\bf k})$ of the diagonal elements, as well as
the non-diagonal elements $n_{\uparrow\downarrow}({\bf k})$ and $n_{\downarrow\uparrow}({\bf k})$,
are of the order of ${\cal O}(1/k^{6})$
in the large-$k$ limit. However, in the presence of SOC, a new term can be induced in the order $1/k^{5}$ for these elements.
This term is
proportional to $C$. Therefore,
the SOC can significantly modify the leading-order behavior of
$\delta n({\bf k})$, $n_{\uparrow\downarrow}({\bf k})$ and $n_{\downarrow\uparrow}({\bf k})$.
\end{itemize}

\textbf{(C)} We find that the adiabatic relation is not modified by
the SOC. Explicitly, when $|\Psi\rangle$ is an eigen-state of the
many-body Hamiltonian $H$, i.e., $H|\Psi\rangle=E|\Psi\rangle$,
then we still have $\frac{\partial E}{\partial(-1/a)}=\frac{\hbar^{2}}{4\pi m}C$,
with $m$ being the single-atom mass.

\textbf{(D)} We derive the energy functional for our system (see Eq.
(\ref{efunal-1})), in which the average energy ${\cal E}=\langle\Psi|H|\Psi\rangle$
of an arbitrary many-body state $|\Psi\rangle$ is expressed in terms
of the single-atom momentum and position distribution functions (i.e.,
$\rho_{\sigma\sigma^{\prime}}({\bf r},{\bf r})$ and $n_{\sigma\sigma^{\prime}}({\bf k})$),
as well as the contact $C$. We find that the SOC induce a new term
for the energy functional, which is the average value of the SOC term
in the Hamiltonian of all the atoms.

Our above results show that in the large-momentum limit the leading-order
behavior of the universal relations are usually not modified by the
SOC. This is can be explained as follows. The SOC is a linear function
of the one-atom momentum ${\bf p}$, while the kinetic energy ${\bf p}^{2}/(2m)$
is a quadric function of ${\bf p}$. Thus, in the large-momentum limit
the SOC is much smaller than ${\bf p}^{2}/(2m)$, and thus cannot
modify the leading-order behavior of the system.

The remainder of this paper is organized as follows. In Sec. II we
describe our model and notations. In Sec. III we derive the short-range
expansion of the single-atom spatial correlation function. $\rho_{\sigma\sigma^{\prime}}({\bf r},{\bf r}+{\bf b})$.
Using this result, in Sec. IV we derive the large-momentum expansion
of the single-atom momentum distribution function $n_{\sigma\sigma^{\prime}}({\bf k})$.
The adiabatic relation and energy functional are derived in In Sec.
V, and a summary and discussion are given in Sec. VI. In the appendix
we present some details of our calculation.

\section{System and wave function}

As shown above, we consider the system with $N$ identical ultracold
SO-coupled pseudo-spin-$1/2$ Fermi atoms. The single-atom Hamiltonian
is given by
\begin{equation}
H_{{\rm 1b}}=\frac{\textbf{p}^{2}}{2m}+\sum_{\alpha,\beta=x,y,z}\lambda_{\alpha\beta}\sigma_{\alpha}p_{\beta}+U(\textbf{r}),\label{h1b}
\end{equation}
with ${\bf p}$ and ${\bf r}$ being the atomic momentum and position,
respectively, and $\sigma_{\alpha}$ ($\alpha=x,y,z$) being the Pauli
operators of the atomic spin. Here $U(\textbf{r})$ is the single-atom
trapping potential, which could be either diagonal or non-diagonal
in the pseudo-spin basis.

The Hilbert space ${\cal H}$ of our system is given by ${\cal H}={\cal H}_{r}\otimes{\cal H}_{s}$,
where ${\cal H}_{r}$ and ${\cal H}_{s}$ are the Hilbert spaces for
the $N$-atom spatial motion and $N$-atom spin, respectively. Here
we use $|\rangle$ to denote the total quantum state in ${\cal H}$,
and use $|\rangle_{r}$ and $|\rangle_{s}$ to denote the states in
${\cal H}_{r}$ and ${\cal H}_{s}$, respectively. In this paper we
work in the ``coordinate representation'' where the $N$-body state
$|\Psi\rangle$ is described by a spinor wave function $|\Psi({\bf r}_{1},...,{\bf r}_{N})\rangle_{s}$
which is defined as
\begin{equation}
|\Psi({\bf r}_{1},...,{\bf r}_{N})\rangle_{s}\equiv\ _{r}\!\langle{\bf r}_{1},...,{\bf r}_{N}|\Psi\rangle,\label{wf}
\end{equation}
with $|{\bf r}_{1},...,{\bf r}_{N}\rangle_{r}$ being the eigen-state
of the position of the atoms. Notice that $|\Psi({\bf r}_{1},...,{\bf r}_{N})\rangle_{s}$
is a $({\bf r}_{1},...,{\bf r}_{N})$-dependent vector in the space
${\cal H}_{s}$.

We further assume that there is an $s$-wave short-range two-body
interaction between atoms in different pseudo-spin state. In the presence
of SOC this interaction can be described by various zero-range models,
e.g., the renormalized delta interaction \cite{Peng soc2} and the
modified Bethe-Peierls boundary condition (MPBC) \cite{Peng soc1}.
In this work we use the MPBC approach. In this approach, the wave
function $|\Psi({\bf r}_{1},...,{\bf r}_{N})\rangle_{s}$ are defined
in the region with ${\bf r}_{i}\neq{\bf r}_{j}$ for $\forall i,j\in(1,...,N)$,
and the total Hamiltonian $H$ of the $N$ atoms is just the summation
of the free-Hamiltonians of each atom, i.e.,
\begin{equation}
H=\sum_{i=1}^{N}H_{{\rm 1b}}^{(i)},\label{h}
\end{equation}
with $H_{{\rm 1b}}^{(i)}$ being the one-body Hamiltonian of the $i$-th
atom. In addition, when two atoms (for instance, atoms 1 and 2) are
close, in our notation the $N$-body wave function should satisfy
\begin{widetext}
\begin{eqnarray}
\lim_{|{\bf r}_{12}|\rightarrow0}|\Psi(\textbf{r}_{1}\cdots,\textbf{r}_{N})\rangle_{s} & = & \left\{ \left(\frac{1}{|\textbf{r}_{12}|}-\frac{1}{a}\right)|S\rangle_{12}-i\left[\frac{\textbf{r}_{12}}{|\textbf{r}_{12}|}\cdot{\bf {\bf G}}\right]|S\rangle_{12}\right\} \otimes|A(\frac{\textbf{r}_{1}+\textbf{r}_{2}}{2};\textbf{r}_{3},\cdots,\textbf{r}_{N})\rangle_{3,\cdots,N}+{\cal O}(|\textbf{r}_{12}|).\nonumber \\
\label{bp}
\end{eqnarray}
\end{widetext}
Here ${\bf r}_{12}={\bf r}_{1}-{\bf r}_{2}$, $a$
is the scattering length, $|S\rangle_{12}=(|\uparrow\rangle_{1}|\downarrow\rangle_{2}-|\downarrow\rangle_{1}|\uparrow\rangle_{2})/\sqrt{2}$
is the singlet spin state of the atoms $1$ and $2$, and ${\bf G}=(G_{x},G_{y},G_{z})$
is a vector-type spin operator with the component $G_{\beta}=\sum_{\alpha=x,y,z}\lambda_{\alpha\beta}(\sigma_{\alpha}^{(1)}-\sigma_{\alpha}^{(2)})/2$
($\beta=x,y,z$), where $\sigma_{x,y,z}^{(i)}$ ($i=1,2$) is the
Pauli operator of the $i$-th atom.
The term with operator ${\bf G}$
describes the SOC-induced modification for the short-range behavior
of the wave function. In Eq. (\ref{bp}) $|A(\frac{\textbf{r}_{1}+\textbf{r}_{2}}{2};\textbf{r}_{3},\cdots,\textbf{r}_{N})\rangle_{3,\cdots,N}$
is a spin state of the atoms $3,...,N$, which depends on the positions
of these atoms as well as the center-of-mass position of the atoms
1 and 2. As shown below, the important parameters in the universal
relations, e.g., the contact $C$, can be expressed in terms of $|A\rangle_{3,...,N}$.

\section{single-atom spatial correlation function}

In this section we derive the expansion of the single-atom spatial
correlation function $\rho_{\sigma,\sigma^{\prime}}({\bf r},{\bf r}+{\bf b})$
($\sigma,\sigma^{\prime}=\uparrow,\downarrow$) in the short-range
limit $b\equiv|{\bf b}|\rightarrow0$. Our calculation is done via
the approach in Ref. \cite{TanPRL2011}.

In the coordinate representation, $\rho_{\sigma,\sigma^{\prime}}({\bf r},{\bf r}+{\bf b})$
($\sigma,\sigma^{\prime}=\uparrow,\downarrow$) corresponding to a
state $|\Psi\rangle$, which is defined in Eq. (\ref{rhorr}), can
be expressed as
\begin{eqnarray}
 &  & \rho_{\sigma\sigma^{\prime}}({\bf r},{\bf r}+{\bf b})\nonumber \\
 & = & N\times{\rm Tr}_{2,...,N}\int D_{2}\left[_{1}\!\langle\sigma^{\prime}|\Psi_{{\bf r}+{\bf b}}\rangle_{s}\langle\Psi_{{\bf r}}|\sigma\rangle_{1}\right]\label{rho1b}
\end{eqnarray}
where $D_{2}=\Pi_{i=2}^{N}d{\bf r}_{i}$,
\begin{eqnarray}
|\Psi_{\textbf{r}}\rangle_{s}\equiv|\Psi(\textbf{r},{\bf r}_{2}\cdots,\textbf{r}_{N})\rangle_{s},
\end{eqnarray}
$|\sigma\rangle_{1}$ ($\sigma=\uparrow,\downarrow$) is the pesudo-spin
states of atom 1, and ${\rm Tr}_{2,...,N}$ is the trace for the spin
states of atoms $2,...,N$. To derive the small-$b$ behavior of $\rho_{\sigma,\sigma^{\prime}}({\bf r},{\bf r}+{\bf b})$,
we divide the integration $\int D_{2}$ in Eq. (\ref{rho1b}) as
\begin{equation}
\int D_{2}=\int_{{\cal R}_{\varepsilon}}D_{2}+\int_{\bar{{\cal R}}_{\varepsilon}}D_{2}.\label{int}
\end{equation}
Here $\varepsilon$ is a small positive distance which satisfies $\varepsilon>b$,
${\cal R}_{\varepsilon}$ is the region with $|{\bf r}-{\bf r}_{i}|>\varepsilon$
for $\forall i=2,...,N$, i.e., the region where the distances between
atom 1 and all the other atoms are larger than $\varepsilon$, and
$\bar{{\cal R}}_{\varepsilon}$ is the complementary of ${\cal R}_{\varepsilon}$
. Since $\varepsilon>b$, in ${\cal R}_{\varepsilon}$ we also have
$|{\bf r}+{\bf b}-{\bf r}_{i}|>0$ for $\forall i=2,...,N$. According
to (\ref{bp}), this fact implies that in the region ${\cal R}_{\varepsilon}$
the wave function $|\Psi_{\textbf{r}+{\bf f}}\rangle_{s}$, as a function
of ${\bf f}$, is analytical when $|{\bf f}|\leqslant|{\bf b}|$.
Thus, in ${\cal R}_{\varepsilon}$ we can expand $|\Psi_{\textbf{r}+{\bf b}}\rangle_{s}$
as a Taylor series of ${\bf b}$:
\begin{eqnarray}
|\Psi_{\textbf{r}+{\bf b}}\rangle_{s} & = & |\Psi_{\textbf{r}}\rangle_{s}+{\bf b}\cdot\nabla_{{\bf r}}|\Psi_{\textbf{r}}\rangle_{s}\nonumber \\
 &  & +\frac{1}{2}\sum_{\alpha,\beta=x,y,z}b_{\alpha}b_{\beta}\frac{\partial^{2}}{\partial r_{\alpha}\partial r_{\beta}}|\Psi_{\textbf{r}}\rangle_{s}+....,\label{taylor}
\end{eqnarray}
where $b_{x,y,z}$ and $r_{x,y,z}$ are the components of ${\bf b}$
and ${\bf r}$, respectively. On the other hand, in $\bar{{\cal R}}_{\varepsilon}$
the wave function $|\Psi_{\textbf{r}+{\bf f}}\rangle_{s}$ may diverges
at some point ${\bf f}_{0}$ which satisfies $|{\bf f}_{0}|<|{\bf b}|$.
Namely, $|{\bf b}|$ may larger than the divergence radius of the
function $|\Psi_{\textbf{r}+{\bf f}}\rangle_{s}$. Thus, in $\bar{{\cal R}}_{\varepsilon}$
we cannot do the expansion (\ref{taylor}) for $|\Psi_{\textbf{r}+{\bf b}}\rangle_{s}$.
Therefore, in Eq. (\ref{rho1b}) we can use the expression (\ref{taylor})
to calculate the integration $\int_{{\cal R}_{\varepsilon}}D_{2}$,
and use Eq. (\ref{bp}) with ${\bf r}_{1}={\bf r}+{\bf b}$ or ${\bf r}_{1}={\bf r}$
to calculate $\int_{\bar{{\cal R}}_{\varepsilon}}D_{2}$. For the
small-$b$ case, with direct calculation we finally expand $\rho_{\sigma,\sigma^{\prime}}({\bf r},{\bf r}+{\bf b})$
as a series of ${\bf b}$ (Appendix \ref{app1}.1):
\begin{eqnarray}
 &  & \rho_{\sigma\sigma^{\prime}}({\bf r},{\bf r}+{\bf b})=\nonumber \\
 &  & \rho_{\sigma\sigma^{\prime}}({\bf r},{\bf r})+\textbf{b}\cdot\textbf{u}_{\sigma\sigma^{\prime}}(\textbf{r})-\frac{b}{8\pi}{\cal C}(\textbf{r})\delta_{\sigma\sigma^{\prime}}\nonumber \\
 &  & +\frac{b^{2}}{24\pi a}{\cal C}(\textbf{r})\delta_{\sigma\sigma^{\prime}}+\frac{1}{2}\sum_{\alpha,\beta=x,y,z}\nu_{\sigma\sigma^{\prime}}^{(\alpha\beta)}(\textbf{r})b_{\alpha}b_{\beta}\nonumber \\
 &  & -\frac{\pi}{2}b\textbf{b}\cdot[3\textbf{w}(\textbf{r})+\textbf{w}^{*}(\textbf{r})]\delta_{\sigma\sigma^{\prime}}
+s_{\sigma,\sigma^\prime}({\bf r},{\bf b})+{\cal O}(b^{3}),\nonumber\\\label{rhor}
\end{eqnarray}
where $\delta_{\sigma\sigma^{\prime}}$ is the Kronecker symbol.

The quantities in each order of the r.h.s of Eq. (\ref{rhor}) are
defined and explained as follows.

\textit{0th-order term:} $\rho_{\sigma\sigma^{\prime}}({\bf r},{\bf r})$
is (up to a global factor) the density matrix element of the atom
at position ${\bf r}$. If the SOC were absent and the trapping potential
$U({\bf r})$ is diagonal in the pseudo-spin basis, we have $\rho_{\uparrow\downarrow}({\bf r},{\bf r})=\rho_{\downarrow\uparrow}(\textbf{r},\textbf{r})=0$.

\textit{1st-order terms:} The factors $\textbf{u}_{\sigma\sigma^{\prime}}(\textbf{r})$
and ${\cal C}({\bf r})$ are defined as
\begin{eqnarray}
 &  & \textbf{u}_{\sigma\sigma^{\prime}}(\textbf{r})=N\times{\rm Tr}_{2,...,N}\nonumber \\
 &  & \lim_{\eta\rightarrow0^{+}}\int_{{\cal R}_{\eta}}D_{2}\left[_{1}\langle\sigma^{\prime}|\left(\nabla_{{\bf r}}|\Psi_{{\bf r}}\rangle_{s}\right){}_{s}\!\langle\Psi_{{\bf r}}|\sigma\rangle_{1}\right],\nonumber
\end{eqnarray}
and
\begin{equation}
{\cal C}(\textbf{r})=8\pi^{2}N(N-1)\int D_{3}\left[_{3,...,N}\!\langle A_{{\bf r}}|A_{{\bf r}}\rangle_{3,...,N}\right],\label{cr}
\end{equation}
respectively, with $D_{3}=\Pi_{i=3}^{N}d{\bf r}_{i}$ and
\begin{eqnarray}
|A_{{\bf r}}\rangle_{3,...,N}\equiv|A({\bf r};\textbf{r}_{3},\cdots,\textbf{r}_{N})\rangle_{3,\cdots,N}.\label{biga}
\end{eqnarray}
Below we will show that the contact $C$ for our systems can be defined
as the integration of the factor ${\cal C}({\bf r})$ with ${\bf r}$.
Thus, ${\cal C}({\bf r})$ can be considered the ``contract density''
of our system.

\textit{2nd order terms:} The factors $\nu_{\sigma\sigma^{\prime}}^{(\alpha\beta)}(\textbf{r})$,
${\bf w}(\textbf{r})$ and $s_{\sigma,\sigma^\prime}({\bf r},{\bf b})$
are given by
\begin{eqnarray}
 &  & \nu_{\sigma\sigma^{\prime}}^{(\alpha\beta)}(\textbf{r})=N\times{\rm Tr}_{2,...,N}\nonumber \\
 &  & \lim_{\eta\rightarrow0^{+}}\int_{{\cal R}_{\eta}}D_{2}\left[_{1}\langle\sigma^{\prime}|\left(\frac{\partial^{2}}{\partial r_{\alpha}\partial r_{\beta}}|\Psi_{{\bf r}}\rangle_{s}\right){}_{s}\!\langle\Psi_{{\bf r}}|\sigma\rangle_{1}\right],\nonumber \\
\label{nu}
\end{eqnarray}
and
\begin{eqnarray}
{\bf w}(\textbf{r}) & = & \frac{1}{2}N(N-1)\int D_{3}\left[_{3,...,N}\!\langle A_{{\bf r}}|\left(\nabla_{{\bf r}}|A_{{\bf r}}\rangle_{3,...,N}\right)\right],\nonumber \\
\label{w}
\end{eqnarray}
and

\begin{eqnarray}
s_{\sigma,\sigma^\prime}({\bf r},{\bf b})= \frac{i{\cal C}(\textbf{r})}{8\pi}b\sum_{\alpha,\beta=x,y,z}\lambda_{\alpha\beta}B_{\sigma\sigma^{\prime}}^{(\alpha)}b_{\beta},\label{sss}
\end{eqnarray}
respectively, with $r_{x,y,z}$ being the components of ${\bf r}$
and
\begin{equation}
B_{\sigma\sigma^{\prime}}^{(\alpha)}={}_{1}\langle\sigma^{\prime}|\sigma_{\alpha}^{(1)}|\sigma\rangle_{1}.\label{sigma}
\end{equation}

In the absence of the SOC, the term $s_{\sigma,\sigma^\prime}({\bf r},{\bf b})$ of Eq. (\ref{rhor}) would disappear from the expansion of $\rho_{\sigma,\sigma^{\prime}}({\bf r},{\bf r}+{\bf b})$,
while all the other terms still remain. Thus, the SOC does not affect
the behavior of $\rho_{\sigma,\sigma^{\prime}}({\bf r},{\bf r}+{\bf b})$
in the 0th and 1st order of ${\bf b}$, and modify the behavior in
the 2nd order by inducing the last term in the r.h.s. of Eq. (\ref{rhor}),
which is an non-analytical term and proportional to the contact density
${\cal C}({\bf r})$ and the SOC intensity $\lambda_{\alpha\beta}$.
Especially, for the non-diagonal elements $\rho_{\uparrow\downarrow}({\bf r},{\bf r}+{\bf b})$
and $\rho_{\downarrow\uparrow}({\bf r},{\bf r}+{\bf b})$ this SOC-induced
term is the lowest-order non-analytical term.

 Furthermore, the appearance of the  term $s_{\sigma,\sigma^\prime}({\bf r},{\bf b})$ in Eq. (\ref{rhor}) can be mathematically interpreted as follows. As shown above, to calculate $ \rho_{\sigma\sigma^{\prime}}({\bf r},{\bf r}+{\bf b})$ via Eq. (\ref{rho1b}) we require to do the integration $\int_{\bar{{\cal R}}_{\varepsilon}}D_{2}$ in the region $\bar{{\cal R}}_{\varepsilon}$. Furthermore, in that region the states $|\Psi_{\textbf{r}}\rangle_{s}$ and $|\Psi_{\textbf{r}+{\bf b}}\rangle_{s}$ in the to-be-integrated function $_{1}\!\langle\sigma^{\prime}|\Psi_{{\bf r}+{\bf b}}\rangle_{s}\langle\Psi_{{\bf r}}|\sigma\rangle_{1}$ should be expanded via the MPBC shown in Eq. (\ref{bp}). The direct calculation shows that $s_{\sigma,\sigma^\prime}({\bf r},{\bf b})$ is lead by the contribution from the SOC-induced term of the MPBC (i.e., the term proportional to ${\bf G}$ in Eq. (\ref{bp})) to the integration $\int_{\bar{{\cal R}}_{\varepsilon}}D_{2}$.

\section{Momentum Distribution Function}

Using our above results on the short-range expansion of $\rho_{\sigma,\sigma^{\prime}}({\bf r},{\bf r}+{\bf b})$,
we can now study the high-momentum behavior of the single-atom momentum
distribution function $n_{\sigma\sigma^{\prime}}({\bf k})$ defined
in (\ref{rhokk}). This factor is related to $\rho_{\sigma\sigma^{\prime}}({\bf r},{\bf r}+{\bf b})$
via
\begin{equation}
n_{\sigma\sigma^{\prime}}({\bf k})=\int d{\bf r}d{\bf b}e^{-i{\bf k}\cdot{\bf b}}\rho_{\sigma\sigma^{\prime}}({\bf r},{\bf r}+{\bf b}).\label{rhok}
\end{equation}
Substituting Eq. (\ref{rhor}) into Eq. (\ref{rhok}), we can obtain
the expression of $n_{\sigma\sigma^{\prime}}({\bf k})$ in the large-$k$
limit (Appendix \ref{app1}.2). In the following we investigate the
SOC-induced effect for the diagonal and non-diagonal elements of the
momentum distribution function.

\subsection{Diagonal elements}

In the large-$k$ limit the diagonal elements $n_{\uparrow\uparrow}({\bf k})$
and $n_{\downarrow\downarrow}({\bf k})$ are given by
\begin{eqnarray}
n_{\sigma\sigma}({\bf k}) & = & \frac{C}{k^{4}}-i\frac{16\pi^{2}}{k^{6}}\textbf{k}\cdot(3\tilde{\textbf{w}}+\tilde{\textbf{w}}^{*})\nonumber \\
 &  & -\eta_{\sigma}\frac{4C}{k^{6}}\sum_{\beta=x,y,z}\lambda_{z\beta}k_{\beta}+{\cal O}\left(\frac{1}{k^{6}}\right),\label{rhokexpan}
\end{eqnarray}
for $\sigma=\uparrow,\downarrow$, with $\eta_{\uparrow}=+1$,
$\eta_{\downarrow}=-1$. Here $k_{x,y,z}$ are the components of ${\bf k}$
and $\tilde{\textbf{w}}=\int\textbf{w}(\textbf{r})d{\bf r}$. In Eq.
(\ref{rhokexpan}) the factor $C$ is defined as
\begin{equation}
C=\int d\textbf{r}{\cal C}(\textbf{r}).\label{contact}
\end{equation}
It is the contact of our system.

Eq. (\ref{rhokexpan}) shows that the leading-order behaviors of $n_{\uparrow\uparrow}({\bf k})$
and $n_{\downarrow\downarrow}({\bf k})$ are not changed by the SOC.
Explicitly, no matter if there is an SOC we always have
\begin{equation}
\lim_{k\rightarrow\infty}n_{\uparrow\uparrow}({\bf k})=\lim_{k\rightarrow\infty}n_{\downarrow\downarrow}({\bf k})=\frac{C}{k^{4}}.\label{lim}
\end{equation}
Nevertheless, in the sub-leading order ($1/k^{5}$) the SOC modify
the behaviors of $n_{\uparrow\uparrow}({\bf k})$ and $n_{\downarrow\downarrow}({\bf k})$
by introducing a new term $\mp\frac{4C}{k^{6}}\sum_{\beta=x,y,z}\lambda_{z\beta}k_{\beta}$,
which is proportional to the SOC intensity $\lambda_{\alpha\beta}$
and the contact $C$.

Furthermore, when we calculate $n_{
\sigma\sigma}({\bf k})$ by substituting Eq. (\ref{rhor}) into Eq. (\ref{rhok}), we can find that the  term $-\eta_{\sigma}\frac{4C}{k^{6}}\sum_{\beta=x,y,z}\lambda_{z\beta}k_{\beta}$ of Eq. (\ref{rhokexpan}) is actually induced by the term $s_{\sigma,\sigma^\prime}({\bf r},{\bf b})$ of Eq. (\ref{rhor}) (Appendix A). In addition, as shown in the end of Sec. III, $s_{\sigma,\sigma^\prime}({\bf r},{\bf b})$ is lead by the SOC-induced term (i.e., the term proportional to ${\bf G}$) of the MBPC Eq. (\ref{bp}), which describes the behavior of the wave function in the short-range limit.
Therefore, the SOC-induced term $-\eta_{\sigma}\frac{4C}{k^{6}}\sum_{\beta=x,y,z}\lambda_{z\beta}k_{\beta}$ in Eq. (\ref{rhokexpan}) can be interpreted as a result of the SOC-induced modification of  the short-range behavior of the wave function.

With the above results we can further get the behavior of the total
momentum distribution $n({\bf k})$, which is given by
\begin{equation}
n({\bf k})=n_{\uparrow\uparrow}({\bf k})+n_{\downarrow\downarrow}({\bf k}).\label{nk-1}
\end{equation}
Substituting Eq. (\ref{rhokexpan}) into Eq. (\ref{nk-1}) we obtain
that in the large-$k$ limit
\begin{equation}
n({\bf k})=\frac{2C}{k^{4}}-i\frac{32\pi^{2}}{k^{6}}{\bf k}\cdot(3\tilde{\textbf{w}}+\tilde{\textbf{w}}^{*})+{\cal O}(\frac{1}{k^{6}}).\label{nkk-1}
\end{equation}
In the absence of the the SOC, both of the two terms in the r.h.s.
of Eq. (\ref{nkk-1}) still exists. Thus, the behavior of $n({\bf k})$
is not modified by the SOC up to the order of $1/k^{5}$.

We can further define the momentum-distribution difference of atoms with spin $\uparrow$ and $\downarrow$ as
\begin{eqnarray}
\delta n({\bf k})=n_{\uparrow\uparrow}({\bf k})-n_{\downarrow\downarrow}({\bf k}).
\end{eqnarray}
Thus, Eq. (\ref{rhokexpan}) yields that
\begin{eqnarray}
\delta n({\bf k})=-\frac{8C}{k^{6}}\sum_{\beta=x,y,z}\lambda_{z\beta}k_{\beta}+{\cal O}\left(\frac{1}{k^{6}}\right).\label{nr}
\end{eqnarray}
This result shows that, if there were no SOC (i.e., $\lambda_{\alpha\beta}=0$),
then $\delta n({\bf k})$
is at most on the order of ${\cal O}(1/k^{6})$, no matter if the
trapping potential $U({\bf r})$ is diagonal or non-diagonal in the
pseudo-spin basis. However, in the presence of the SOC, a term $-\frac{8C}{k^{6}}\sum_{\beta=x,y,z}\lambda_{z\beta}k_{\beta}$
can be induced in the order of $1/k^{5}$. Namely, the SOC can significantly
modify the behavior of $\delta n({\bf k})$ by changing the leading order
from at most $1/k^{6}$ to $1/k^{5}$.
This result also implies that the
SOC-induced modification of the single-atom momentum distribution may be experimentally detected via the measurement of the large-$k$ behavior of $\delta n({\bf k})$.

\subsection{Non-diagonal elements}

The situation is quite different for the non-diagonal elements $n_{\uparrow\downarrow}({\bf k})$
and $n_{\downarrow\uparrow}({\bf k})$. Substituting Eq. (\ref{rhor})
into Eq. (\ref{rhok}), we find that in the large-$k$ limit we have

\begin{equation}
n_{\uparrow\downarrow}({\bf k})=-\frac{4C}{k^{6}}\sum_{\beta=x,y,z}\left(\lambda_{x\beta}+i\lambda_{y\beta}\right)k_{\beta}+{\cal O}\left(\frac{1}{k^{6}}\right),\label{rhoud}
\end{equation}
and $n_{\downarrow\uparrow}({\bf k})=n_{\uparrow\downarrow}({\bf k})^{\ast}$.
Therefore, similar as $\delta n({\bf k})$, the large-$k$ behaviors of
$n_{\uparrow\downarrow}({\bf k})$ and $n_{\downarrow\uparrow}({\bf k})$
are also significantly modified by the SOC and the leading order
of these terms
 is changed   from at most $1/k^{6}$ to $1/k^{5}$.

\section{Many-Body Energy}

In this section we use our above results to study the relations beween
the many-body energy and the contact, and derive the adiabatic relation
and energy functional for our SO-coupled system.

\subsection{Adiabatic relation}

We consider the case that $|\Psi\rangle$ is an eigen-state (e.g.,
the ground state) of the Hamiltonian $H$, which satisfies the eigen-equaiton
$H|\Psi\rangle=E|\Psi\rangle$ as well as the MBPC (\ref{bp}) with
scattering length $a$. Thus, the eigen-energy $E$ depends on the
value of $a$. Now we derive the adiabatic relation which connects
$\partial E/\partial(-1/a)$ and the contact $C$ for our SO-coupled
system, with the approach shown in Ref. \cite{CastinPRA2012}. To
this end we consider two systems $\alpha$ and $\beta$ with scattering
lengths $a_{\alpha}$ and $a_{\beta}$, respectively, and assume $|\Psi^{(n)}\rangle$
($n=\alpha,\beta$) is the normalized eigen-state of $H$ for system
$n$, with corresponding eigen-energy $E_{n}$. We further assume
these two eigen-energies have the same rank for each system. For instance,
$|\Psi^{(\alpha)}\rangle$ and $|\Psi^{(\beta)}\rangle$ are the ground
states of the systems $\alpha$ and $\beta$, respectively, and $E_{\alpha,\beta}$
are the ground-state energies. Therefore, we have \cite{ip}
\begin{equation}
\lim_{a_{\alpha}\rightarrow a_{\beta}}E_{\alpha}=E_{\beta};\ \ \lim_{a_{\alpha}\rightarrow a_{\beta}}\langle\Psi^{(\alpha)}|\Psi^{(\beta)}\rangle=1.\label{limm}
\end{equation}
On the other hand, with the direct generalization of the approach
used in Ref. \cite{CastinPRA2012}, we can prove that (Appendix \ref{ar})
\begin{eqnarray}
 &  & \frac{E_{\beta}-E_{\alpha}}{\left(\frac{1}{a_{\alpha}}-\frac{1}{a_{\beta}}\right)}\langle\Psi^{(\alpha)}|\Psi^{(\beta)}\rangle\nonumber \\
 & = & \frac{2\pi\hbar^{2}}{m}N(N-1)\int d\textbf{r}\int D_{3}\left[_{3,...,N}\!\langle A_{{\bf r}}^{(\alpha)}|A_{{\bf r}}^{(\beta)}\rangle_{3,...,N}\right].\nonumber \\
\label{ad1}
\end{eqnarray}
Here $|A_{{\bf r}}^{(n\rangle}\rangle_{3,...,N}$ ($n=\alpha,\ \beta$)
is the state $|A({\bf r};\textbf{r}_{3},\cdots,\textbf{r}_{N})\rangle_{3,\cdots,N}$
of system $n$, and thus satisfies $\lim_{a_{\alpha}\rightarrow a_{\beta}}|A_{{\bf r}}^{(\alpha\rangle}\rangle_{3,...,N}=|A_{{\bf r}}^{(\beta\rangle}\rangle_{3,...,N}$.
Substituting taking the limit $a_{\alpha}\rightarrow a_{\beta}$ for
Eq. (\ref{ad1}) and using Eqs. (\ref{limm}, \ref{cr}, \ref{contact}),
we obtain the adiabatic relation
\begin{equation}
\frac{\partial E}{\partial\left(-1/a\right)}=\frac{\hbar^{2}C}{4\pi m},\label{arr}
\end{equation}
which has the same form as in the systems without SOC.

\subsection{Energy functional}

If $|\Psi\rangle$ is an arbitrary many-body state, the average energy
can be defined as ${\cal E}=\langle\Psi|H|\Psi\rangle$ and can be
expressed as a functional of the single-atom momentum and position
distribution function. Now we derive the expression of this energy
functional, with the help of our previous result in Eq. (\ref{rhor}).
Here we use the similar approach as Ref. \cite{TanPRL2011}. We first
define a function $J(\beta)$ as
\begin{equation}
J(\beta)=\frac{1}{(2\pi)^{3}}\sum_{\sigma=\uparrow,\downarrow}\int d\textbf{k}n_{\sigma\sigma}({\bf k})e^{-\beta\frac{\hbar^{2}k^{2}}{2m}}.\label{j0}
\end{equation}
Substituting Eq. (\ref{rhok}) into this definition, we find that
$J(\beta)$ can be re-expressed as
\begin{equation}
J(\beta)=\sum_{\sigma=\uparrow,\downarrow}\int d\textbf{r}d{\bf b}U_{\beta}({\bf b})\rho_{\sigma\sigma}(\textbf{r},\textbf{r}+{\bf b}),\label{j}
\end{equation}
where $U_{\beta}({\bf b})=(2\pi\beta\hbar^{2}/m)^{-\frac{3}{2}}e^{-\frac{m}{2\beta\hbar^{2}}|{\bf b}|^{2}}$.
Using Eq. (\ref{rhor}), we further obtain
\begin{eqnarray}
J(\beta) & = & N-\frac{C\hbar\sqrt{\beta}}{\sqrt{2m}\pi^{\frac{3}{2}}}+\frac{C\beta\hbar^{2}}{4\pi ma}+\beta K+{\cal O}(\beta^{3/2}),\nonumber \\
\label{jj}
\end{eqnarray}
with $N=\int d{\bf r}\left[\rho_{\uparrow\uparrow}(\textbf{r},\textbf{r})+\rho_{\downarrow\downarrow}(\textbf{r},\textbf{r})\right]$
being the total atom number and $K=\frac{\hbar^{2}}{2m}\sum_{\substack{\alpha=x,y,z,\\
\sigma=\uparrow,\downarrow
}
}\int d\textbf{r}\nu_{\sigma\sigma}^{(\alpha\alpha)}(\textbf{r})$. On the other hand, the definition ${\cal E}=\langle\Psi|H|\Psi\rangle$
of the average energy yields
\begin{eqnarray}
{\cal E} & = & \int d\textbf{r}\bar{U}({\bf r})-K+\sum_{\substack{\alpha,\beta=x,y,z,\\
\sigma,\sigma^{\prime}=\uparrow,\downarrow
}
}\int d{\bf k}\left[\lambda_{\alpha\beta}B_{\sigma\sigma^{\prime}}^{(\alpha)}k_{\beta}
n_{\sigma^{\prime}\sigma}({\bf k})\right],\nonumber \\
\label{22}
\end{eqnarray}
where $B_{\sigma\sigma^{\prime}}^{(x,y,z)}$ are the matrix-elements
of the Pauli operator, as defined in Eq. (\ref{sigma}), and $\bar{U}({\bf r})=\sum_{\sigma,\sigma^{\prime}=\uparrow,\downarrow}U_{\sigma\sigma^{\prime}}(\textbf{r})d_{\sigma\sigma^{\prime}}(\textbf{r})$,
with $U_{\sigma\sigma^{\prime}}(\textbf{r})$ being the matrix-element
of the trapping potential in the one-body pseudo-spin basis
 and
\begin{eqnarray}
d_{\sigma\sigma^{\prime}}({\bf r})=\rho_{\sigma\sigma^{\prime}}({\bf r},{\bf r}).
\end{eqnarray}Substituting
Eq. (\ref{22}) into Eq. (\ref{jj}) and using Eq. (\ref{j0}) as
well as the relation
\begin{equation}
\frac{\hbar}{2\pi^{\frac{3}{2}}\sqrt{2m\beta}}=\frac{1}{(2\pi)^{3}}\int d{\bf k}\left(\frac{\hbar^{2}}{mk^{2}}e^{-\beta\frac{\hbar k^{2}}{2m}}\right),\label{rell}
\end{equation}
we find that
\begin{eqnarray}
 &  & \frac{1}{(2\pi)^{3}}\sum_{\sigma=\uparrow,\downarrow}\int d\textbf{k}e^{-\beta\frac{\hbar k^{2}}{2m}}\left[n_{\sigma\sigma}({\bf k})-\frac{C}{k^{4}}\right]\nonumber \\
 & = & N+\frac{C\beta\hbar^{2}}{4\pi am}-\beta\left[{\cal E}-\int d{\bf r}\bar{U}({\bf r})\right]\nonumber \\
 &  & +\beta\sum_{\substack{\alpha,\beta=x,y,z,\\
\sigma,\sigma^{\prime}=\uparrow,\downarrow
}
}\int d{\bf k}\left[\lambda_{\alpha\beta}B_{\sigma\sigma^{\prime}}^{(\alpha)}k_{\beta}n_{\sigma^{\prime}\sigma}({\bf k})\right]+{\cal O}(\beta^{3/2}).\nonumber \\
\label{ffor}
\end{eqnarray}
Doing the operation $\left.\frac{d}{d\beta}(...)\right|_{\beta=0}$
for both of the two sides of Eq. (\ref{ffor}), we finally obtain
the expression of the energy functional:
\begin{eqnarray}
{\cal E} & = & \frac{\hbar^{2}C}{4\pi am}+\frac{1}{(2\pi)^{3}}\sum_{\sigma=\uparrow,\downarrow}\int d\textbf{k}\frac{\hbar k^{2}}{2m}\left[n_{\sigma\sigma}({\bf k})-\frac{C}{k^{4}}\right]\nonumber \\
 &  & +\sum_{\sigma,\sigma^{\prime}=\uparrow,\downarrow}\int d\textbf{r}\left[U_{\sigma\sigma^{\prime}}(\textbf{r}) d_{\sigma\sigma^{\prime}}(\textbf{r})\right]\nonumber \\
 &  & +\frac{1}{(2\pi)^{3}}\sum_{\substack{\alpha,\beta=x,y,z,\\
\sigma,\sigma^{\prime}=\uparrow,\downarrow
}
}\int d{\bf k}\left[\lambda_{\alpha\beta}B_{\sigma\sigma^{\prime}}^{(\alpha)}k_{\beta}n_{\sigma^{\prime}\sigma}({\bf k})\right],\nonumber \\
\label{efunal-1}
\end{eqnarray}
This result shows that the SOC modify the energy functional by introducing
the last term of Eq. (\ref{efunal-1}), which is just the average
value of the SOC term in the total Hamiltonian $H$ on state $|\Psi\rangle$.
Thus, this energy functional is the direct generalization of the one
for the systems without SOC.

\section{Summary}

In this paper we derive the universal relations for ultracold two-component
Fermi gases with arbitrary type SOC. We obtain the short-range and
high-momentum expansions of the single-atom spatial correlation function
$\rho_{\sigma\sigma^{\prime}}({\bf r},{\bf r}+{\bf b})$ ($\sigma,\sigma^{\prime}=\uparrow,\downarrow$)
and momentum distribution function $n_{\sigma\sigma^{\prime}}({\bf k})$.
We find that the SOC significantly modify the leading-order behaviors
of $n_{\uparrow\downarrow}({\bf k})$,
$n_{\downarrow\uparrow}({\bf k})$ and the distribution difference $n_{\uparrow\uparrow}({\bf k})-n_{\downarrow\downarrow}({\bf k})$,
and modify the sub-leading-order behaviors of each diagonal element $n_{\uparrow\uparrow}({\bf k})$ and
$n_{\downarrow\downarrow}({\bf k})$ as well as all the four elements
of $\rho_{\sigma\sigma^{\prime}}({\bf r},{\bf r}+{\bf b})$. All of these modifications are proportional to
the contact $C$, and are indued by the SOC-induced term in the MBPC
(\ref{bp}), i.e., the term proportional to the operator ${\bf G}$. We
further derive the adiabatic relation and energy functional for our
system. Our calculations can be generalized to the two-dimensional systems.

Our results are helpful for the further theoretical and experimental
studies of ultracold gases with SOC or the ultracold gases where the single-atom
trapping potential $U({\bf r})$ is non-diagonal in the pseudo-spin
basis. We should also be careful that, the expressions of the SOC
intensity $\lambda_{\alpha\beta}$ ($\alpha,\beta=x,y,z$) and the
potential $U({\bf r})$ are ``frame-dependent\char`\"{}. For instance,
for the experimental systems with one-dimensional SOC induced by the
Raman laser beams, in the lab frame we have $\lambda_{\alpha\beta}=0$
for all $\alpha,\beta$ and $U({\bf r})=\frac{\Omega}{2}e^{ik_{{\rm L}}x}|\uparrow\rangle\langle\downarrow|+h.c.$,
while in the rotated frame we have $U({\bf r})=\frac{\Omega}{2}|\uparrow\rangle\langle\downarrow|+h.c.$
and $\lambda_{\alpha\beta}=(\hbar k_{{\rm L}}/2)\delta_{\alpha,z}\delta_{\beta,x}$.
Here $\Omega$ and $k_{{\rm L}}$ being the effective Rabi frequency
and wave vector of the Raman beams, respectively. Therefore, when
working in a certain frame we should correctly use the corresponding
expressions of $\lambda_{\alpha\beta}$ and the potential $U({\bf r})$.
\begin{acknowledgments}
We thank Shina Tan for helpful discussions. We also thank
the referees for improving the quality of this manuscript. This work has been supported
by the Natural Science Foundation of China under Grant Nos. 11434011
and 11674393, the Fundamental Research Funds for the Central Universities,
and the Research Funds of Renmin University of China under Grant No.
16XNLQ03 and No. 17XNH054.

\bigskip{}

\end{acknowledgments}

\textit{Note added.}-Recently, we became aware of a parallel paper
by P. Zhang and N. Sun \cite{Zhangpengfei2018}, where the large-$k$ expansion of
the single-atom momentum distribution and the adiabatic relation are derived for the
systems with one-dimensional and three-dimensional isotropic SOC and $U({\bf r})=0$ via the effective-field theory approach. Our
 results are consistent.

\appendix
%dummy comment inserted by tex2lyx to ensure that this paragraph is not empty%dummy comment inserted by tex2lyx to ensure that this paragraph is not empty%dummy comment inserted by tex2lyx to ensure that this paragraph is not empty%dummy comment inserted by tex2lyx to ensure that this paragraph is not empty%dummy comment inserted by tex2lyx to ensure that this paragraph is not empty%dummy comment inserted by tex2lyx to ensure that this paragraph is not empty

\section{Integrations}

\label{app1}

In this appendix we given some remarks on the calculation of the integration
$\int D_{2}=\int_{{\cal R}_{\varepsilon}}D_{2}+\int_{\bar{{\cal R}}_{\varepsilon}}D_{2}$
for Eq. (\ref{rhor}) and the the integration $\int d{\bf b}$ in
Eq. (\ref{rhok}).

For the integration $\int D_{2}$, we have the following remarks:

(1) To do the small-$b$ expansion, here we only need to consider
the cases where both $b$ and $\varepsilon$ are very small. On the
other hand, in principle, in the region $\bar{{\cal R}}_{\varepsilon}$
it is possible that more than one atoms are in the region with center
${\bf r}$ and radius $\varepsilon$. For instance, both atom 2 and
atom 3 are in this region. Since in the wave function $|\Psi_{{\bf r}}\rangle$
the vector ${\bf r}$ is the position of atom 1, in this example the
three atoms 1, 2 and 3 are very close to each other. Due to the Pauli
principle, this possibility is very small and here we ignore this
possibility. Namely, we assume in $\bar{{\cal R}}_{\varepsilon}$
only one atom is in the region with center ${\bf r}$ and radius $\varepsilon$.

(2) In addition, since the finial result of the integration $\int D_{2}=\int_{{\cal R}_{\varepsilon}}D_{2}+\int_{\bar{{\cal R}}_{\varepsilon}}D_{2}$
is independent of the value of $\varepsilon$, here we calculate $\int_{{\cal R}_{\varepsilon}}D_{2}$
and $\int_{\bar{{\cal R}}_{\varepsilon}}D_{2}$ to the 0th power of
$\varepsilon$ and then sum the results.

(3) Furthermore, also due to the identity of Fermonic atoms, the to-be-integrated
function ${\rm Tr}_{2,...,N}\left[_{1}\!\langle\sigma^{\prime}|\Psi_{{\bf r}+{\bf b}}\rangle_{s}\langle\Psi_{{\bf r}}|\sigma\rangle_{1}\right]$
is invariable under the exchange of ${\bf r}_{i}$ and ${\bf r}_{j}$
for $\forall i,j\in(2,...,N)$. Thus, in our calculation we have
\begin{equation}
\int_{\bar{{\cal R}}_{\varepsilon}}D_{2}=(N-1)\int_{\bar{{\cal R}}_{\varepsilon}^{(2)}}D_{2},\label{int2}
\end{equation}
with $\bar{{\cal R}}_{\varepsilon}^{(2)}$ being the region with $|{\bf r}_{2}-{\bf r}|<\varepsilon$.

For the integration $\int d{\bf b}$ in Eq. (\ref{rhok}) we have
the following remarks:

(1) When we do this integration we first replace $\int d{\bf b}e^{-i{\bf k}\cdot{\bf b}}\rho_{\sigma\sigma^{\prime}}({\bf r},{\bf r}+{\bf b})$
with $\lim_{\zeta\rightarrow0^{+}}\int d{\bf b}e^{-i{\bf k}\cdot{\bf b}}e^{-\zeta b}\rho_{\sigma\sigma^{\prime}}({\bf r},{\bf r}+{\bf b}),$
and then replace $\rho_{\sigma\sigma^{\prime}}({\bf r},{\bf r}+{\bf b})$
with the expansion (\ref{rhor}).

(2) In the large-$k$ limit, the contributions from the analytical
terms of Eq. (\ref{rhor}) to the Fourier transformation $\int d{\bf b}e^{-i{\bf k}\cdot{\bf b}}$
would exponentially decay with $k$. Since now we want to extract
the power terms which behave as $1/k^{n}$ with $n$ an integer, we
only need to consider the contributions from the non-analytical terms
of Eq. (\ref{rhor}), i.e., the terms proportional to $b$, $b^{2}$,
$b{\bf b}$ and $bb_{\beta}$. For instance, the direct calculation gives
\begin{equation}
\lim_{\zeta\rightarrow0^{+}}\int d{\bf b}e^{-i{\bf k}\cdot{\bf b}}e^{-\zeta b}b=-\frac{8\pi}{k^{4}}.\label{a2-1}
\end{equation}
Thus, we have
\begin{eqnarray}
\lim_{\zeta\rightarrow0^{+}}\int d{\bf b}e^{-i{\bf k}\cdot{\bf b}}e^{-\zeta b}\left[-\frac{b}{8\pi}{\cal C}(\textbf{r})\delta_{\sigma\sigma^{\prime}}\right]=\frac{\cal C({\bf r})}{k^{4}}\delta_{\sigma\sigma^{\prime}}.\nonumber\\
\end{eqnarray}
Similarly, we also have
\begin{eqnarray}
\lim_{\zeta\rightarrow0^{+}}\int d{\bf b}e^{-i{\bf k}\cdot{\bf b}}e^{-\zeta b}s_{\sigma,\sigma}({\bf r},{\bf b})=-\eta_\sigma \frac{4{\cal C}({\bf r})}{k^{6}}\sum_{\beta=x,y,z}\lambda_{z\beta}k_{\beta},\nonumber\\
\end{eqnarray}
for $\sigma=\uparrow,\downarrow$, where $s_{\sigma,\sigma}({\bf r},{\bf b})$ is defined in Eq. (\ref{sss}) and $\eta_{\uparrow}=+1$ and $\eta_{\downarrow}=-1$. This result implies that the term $\eta_\sigma q({\bf k})$ in Eq. (\ref{rhokexpan}) is lead by the  term $s_{\sigma,\sigma}({\bf r},{\bf b})$ of Eq. (\ref{rhor}). The contribution from other non-analytical terms can be derived with
similar calculations.

\begin{widetext}

\section{ Proof of Eq. (\ref{ad1})}

\label{ar} In this appendix we prove Eq. (\ref{ad1}) with the method
in Ref. \cite{CastinPRA2012}. Our following calculation is a generalization
of the calculation in Appendix B of Ref. \cite{CastinPRA2012} to
our case with SOC. As shown in Sec. V.A of our main text, we assume
$|\Psi^{(\alpha)}\rangle$ and $|\Psi^{(\beta)}\rangle$ are eigen-states
of the many-body Hamiltonian $H$ of systems with scattering lengths
$a_{\alpha}$ and $a_{\beta}$, and $E_{\alpha,\beta}$ are the corresponding
eigen-energies. As in the main text, here we define the corresponding
wave functions $|\psi^{(i)}\rangle_{s}\equiv\ _{s}\!\langle{\bf r}_{1},...,{\bf r}_{N}|\Psi^{(i)}\rangle$
$(i=\alpha,\beta)$.

When $a_{\alpha}\neq a_{\beta}$, since $|\Psi^{(\alpha)}\rangle$
and $|\Psi^{(\beta)}\rangle$ satisfy the MBPC for different scattering
lengths, in principle they are in different Hilbert spaces and thus
the inner product between them is not well-defined. Nevertheless,
here for our system, even if $a_{\alpha}\neq a_{\beta},$ we can still
do the integration
\[
I_{\alpha\beta}\equiv\int D_{1}\left(_{s}\!\langle\psi^{(\alpha)}|\psi^{(\beta)}\rangle_{s}\right)=\int D_{1}\left(\langle\Psi^{(\alpha)}|{\bf r}_{1},...,{\bf r}_{N}\rangle_{s}\langle{\bf r}_{1},...,{\bf r}_{N}|\Psi^{(\beta)}\rangle\right),
\]
where $D_{1}=\Pi_{j=1}^{N}d{\bf r}_{j}$, and the integration is defined
in the region where $r_{ij}\equiv|{\bf r}_{i}-{\bf r}_{j}|\neq0$
($i,j=1,...,N$), as in the main text. Thus, we can define the symbol
$\langle\Psi^{(\alpha)}|\Psi^{(\beta)}\rangle$ as $\langle\Psi^{(\alpha)}|\Psi^{(\beta)}\rangle\equiv I_{\alpha\beta}$
for both the cases with $a_{\alpha}=a_{\beta}$ and the cases with
$a_{\alpha}\neq a_{\beta}$. It is clear that when $a_{\alpha}=a_{\beta}$
this definition returns to the usual definition of inner product.

Under this definition, we have
\begin{eqnarray}
(E_{\beta}-E_{\alpha})\langle\Psi^{(\alpha)}|\Psi^{(\beta)}\rangle & = & \langle\Psi^{(\alpha)}|H\Psi^{(\beta)}\rangle-\langle H\Psi^{(\alpha)}|\Psi^{(\beta)}\rangle\nonumber \\
 & = & N\left(\langle\Psi^{(\alpha)}|H_{{\rm 1b}}^{(1)}\Psi^{(\beta)}\rangle-\langle H_{{\rm 1b}}^{(1)}\Psi^{(\alpha)}|\Psi^{(\beta)}\rangle\right).\label{a1}
\end{eqnarray}
Here in the last step we have used Eq. (\ref{h}). Using the expression
(\ref{h1b}) of the one-body Hamiltonian, we find that in our coordinate
representation we have
\begin{eqnarray}
\langle\Psi^{(\alpha)}|H_{{\rm 1b}}^{(1)}\Psi^{(\beta)}\rangle-\langle H_{{\rm 1b}}^{(1)}\Psi^{(\alpha)}|\Psi^{(\beta)}\rangle & = & -\frac{\hbar^{2}}{2m}\int D_{1}\left[_{s}\!\langle\psi^{(\alpha)}|\left(\nabla_{{\bf r}_{1}}^{2}|\psi^{(\beta)}\rangle_{s}\right)-{}_{s}\!\langle\nabla_{{\bf r}_{1}}^{2}\psi^{(\alpha)}|\psi^{(\beta)}\rangle_{s}\right]\nonumber \\
 &  & -i\sum_{\eta,\xi=x,y,z}\lambda_{\alpha\beta}\int D_{1}\left[_{s}\!\langle\psi^{(\alpha)}|\sigma_{\eta}^{(1)}\left(\frac{\partial}{\partial\xi_{1}}|\psi^{(\beta)}\rangle_{s}\right)+{}_{s}\!\langle\frac{\partial}{\partial\xi_{1}}\psi^{(\alpha)}|\sigma_{\eta}^{(1)}|\psi^{(\beta)}\rangle_{s}\right],\label{a2}
\end{eqnarray}
where $\xi_{1}$ ($\xi=x,y,z$) are the three components of ${\bf r}_{1}$,
while $_{s}\!\langle\nabla_{{\bf r}_{1}}^{2}\psi^{(\alpha)}|$ and
$_{s}\!\langle\frac{\partial}{\partial\xi_{1}}\psi^{(\alpha)}|$ are
the Dirac bra corresponding to $\nabla_{{\bf r}_{1}}^{2}|\psi^{(\alpha)}\rangle_{s}$
and $\frac{\partial}{\partial\xi_{1}}|\psi^{(\alpha)}\rangle_{s}$,
respectively. Notice that the integration is done in the region where
$r_{ij}\equiv|{\bf r}_{i}-{\bf r}_{j}|\neq0$ ($i,j=1,...,N$). Using
the divergence theorem, we have
\begin{eqnarray}
\langle\Psi^{(\alpha)}|H_{{\rm 1b}}^{(1)}\Psi^{(\beta)}\rangle-\langle H_{{\rm 1b}}^{(1)}\Psi^{(\alpha)}|\Psi^{(\beta)}\rangle & = & -\int D_{2}\lim_{\epsilon\to0}\int_{\{\forall j\neq1,r_{1j}>\epsilon\}}d{\bf r}_{1}\left(\nabla_{{\bf r}_{1}}\cdot{\bf I}\right),\nonumber \\
 & = & \int D_{2}\lim_{\epsilon\to0}\sum_{j\neq1}\varoiint_{S_{\epsilon}({\bf r}_{j})}{\bf I}\cdot\mathbf{dS},\label{iie}
\end{eqnarray}
with $D_{2}=\Pi_{j=2}^{N}d{\bf r}_{j}$ as we have defined before,
and the surface integral $\varoiint_{S_{\epsilon}({\bf r}_{j})}$
($j=2,...,N$) is done for ${\bf r}_{1}$ in the sphere $S_{\epsilon}({\bf r}_{j})$
which is centered at ${\bf r}_{j}$ and has radius $\epsilon$. Here
the vector area $\mathbf{dS}$ points out of this sphere. In Eq. (\ref{iie})
the vector function ${\bf I}$ is given by
\begin{equation}
{\bf I}=\frac{\hbar^{2}}{2m}\left[_{s}\!\langle\psi^{(\alpha)}|\left(\nabla_{{\bf r}_{1}}|\psi^{(\beta)}\rangle_{s}\right)-{}_{s}\!\langle\nabla_{{\bf r}_{1}}\psi^{(\alpha)}|\psi^{(\beta)}\rangle_{s}\right]+i\left(_{s}\!\langle\psi^{(\alpha)}|{\bf g}|\psi^{(\beta)}\rangle_{s}+{}_{s}\!\langle\psi^{(\alpha)}|{\bf g}|\psi^{(\beta)}\rangle_{s}\right),\label{ii}
\end{equation}
with the spin operator ${\bf g}$ being defined as ${\bf g}=\sum_{\alpha,\beta=x,y,z}\lambda_{\alpha\beta}\sigma_{\alpha}^{(1)}{\bf e}_{\beta}$,
while ${\bf e}_{x,y,z}$ being the unit vector along the three coordinate
axises. In addition, due to the identity of Fermi atoms, the function
${\bf I}$ is invariable under the exchange of ${\bf r}_{l}$ and
${\bf r}_{j}$ ($j,l\neq1$). Therefore, the integration $\varoiint_{S_{\epsilon}({\bf r}_{j})}{\bf I}\cdot\mathbf{dS}$
is same for different $j$, and thus Eq. (\ref{iie}) becomes
\begin{equation}
\langle\Psi^{(\alpha)}|H_{{\rm 1b}}^{(1)}\Psi^{(\beta)}\rangle-\langle H_{{\rm 1b}}^{(1)}\Psi^{(\alpha)}|\Psi^{(\beta)}\rangle=(N-1)\int D_{2}\lim_{\epsilon\to0}\varoiint_{S_{\epsilon}({\bf r}_{2})}{\bf I}\cdot\mathbf{dS}.\label{iie2}
\end{equation}
In addition, in the surphase $S_{\epsilon}({\bf r}_{2})$, we can
express the wave functions $|\psi^{(n)}\rangle_{s}$ ($n=\alpha,\beta$)
via the MBPC (\ref{bp}). Now we define a vector ${\bf u}$ as ${\bf u}\equiv({\bf r}_{1}-{\bf r}_{2})/r_{12}$.
Thus, in surphace $S_{\epsilon}({\bf r}_{2})$ we have $\frac{{\bf r}_{1}+{\bf r}_{2}}{2}={\bf r}_{2}+\epsilon{\bf u}/2$
and
\begin{eqnarray}
|\psi^{(n)}\rangle_{s} & = & \left.\left[\frac{1}{\epsilon}-\frac{1}{a_{n}}-i{\bf u}\cdot{\bf G}+\frac{{\bf u}}{2}\cdot\nabla_{{\bf r}}\right]|S\rangle_{12}\otimes|A_{{\bf r}}^{(n)}\rangle_{3,\cdots,N}\right|_{{\bf r}={\bf r}_{2}}+{\cal O}(\epsilon);\label{s1}\\
\nabla_{{\bf r}_{1}}|\psi^{(n)}\rangle_{s} & = & \left.\left[-\frac{{\bf u}}{\epsilon^{2}}-\frac{1}{2\epsilon}{\bf u}\left({\bf u}\cdot\nabla_{{\bf r}}\right)+\frac{1}{2\epsilon}\nabla_{{\bf r}}\right]|S\rangle_{12}\otimes|A_{{\bf r}}^{(n)}\rangle_{3,\cdots,N}\right|_{{\bf r}={\bf r}_{2}}+{\cal O}(1),\label{s2}\\
 &  & ({\rm for}\ \ n=\alpha,\beta),\nonumber
\end{eqnarray}
where $|A_{{\bf r}}^{(n\rangle}\rangle_{3,...,N}$ ($n=\alpha,\beta$)
is the state $|A({\bf r};\textbf{r}_{3},\cdots,\textbf{r}_{N})\rangle_{3,\cdots,N}$
defined in Eq. (\ref{bp}) for the system with scattering length $a_{n}$.
Substituting Eqs. (\ref{s1}) and (\ref{s2}) into Eq. (\ref{ii})
and then into Eq. (\ref{iie2}), we obtain
\begin{equation}
\langle\Psi^{(\alpha)}|H_{{\rm 1b}}^{(1)}\Psi^{(\beta)}\rangle-\langle H_{{\rm 1b}}^{(1)}\Psi^{(\alpha)}|\Psi^{(\beta)}\rangle=\frac{(N-1)(2\pi)\hbar^{2}}{m}\left(\frac{1}{a_{\alpha}}-\frac{1}{a_{\beta}}\right)\int d\textbf{r}\int D_{3}\left[_{3,...,N}\!\langle A_{{\bf r}}^{(\alpha)}|A_{{\bf r}}^{(\beta)}\rangle_{3,...,N}\right].\label{ir1}
\end{equation}
Further substituting this result into Eq. (\ref{a1}), we obtain Eq.
(\ref{ad1}) in Sec. V.

\end{widetext} %\newpage{}


\begin{thebibliography}{10}
\bibitem{ip} The mathematical definition of $\langle\Psi^{(\alpha)}|\Psi^{(\beta)}\rangle$
for the case with $a_{\alpha}\neq a_{\beta}$ is given in Appendix
\ref{ar}.
\end{thebibliography}
\end{document}